\documentclass[aps,twocolumn]{revtex4}

\usepackage{hyperref}
\usepackage{graphicx}
\usepackage[utf8]{inputenc}
\DeclareGraphicsExtensions{.png,.jpg,.eps,.pdf}
\usepackage{amsmath}
\usepackage{comment}
\usepackage{bm}
\usepackage{dsfont}

\usepackage{xcolor}

\begin{document}

\title{Hubbard model for spin-1 Haldane chains}

\author{
G. Catarina$^{1,2,3,}$\footnote{goncalo.catarina@inl.int}, J. Fern\'{a}ndez-Rossier$^{1,}$\footnote{On leave from Departamento de F\'{i}sica Aplicada, Universidad de Alicante, 03690 San Vicente del Raspeig, Spain.}
}
\affiliation{
$^1$QuantaLab, International Iberian Nanotechnology Laboratory (INL), 4715-330 Braga, Portugal
}
\affiliation{
$^2$Departamento de F\'{i}sica Aplicada, Universidad de Alicante, 03690 San Vicente del Raspeig, Spain
}
\affiliation{
$^3$Centro de F\'{i}sica das Universidades do Minho e do Porto, Universidade do Minho, Campus de Gualtar, 4710-057 Braga, Portugal
}

\date{\today}

%%%%%%%%%%%%%%%%%%%%%%%%%%%%%%%%%%%%%%%%%%%%%%%%%%%%%%%%%%%%%%%%%%%%%%%%%%%%%
%Abstract

\begin{abstract}
The Haldane phase for antiferromagnetic spin-1 chains is a celebrated topological state of matter, featuring gapped excitations and fractional spin-1/2 edge states.
Here, we provide numerical evidence that this phase can be realized with a Hubbard model at half filling, where each $s=1$ spin is stored in a four-site fermionic structure.
We find that the noninteracting limit of our proposed model describes a one-dimensional topological insulator, and we conjecture it to be adiabatically connected to the Haldane phase.
Our work suggests a route to build spin-1 networks using Hubbard model quantum simulators.
\end{abstract}

\maketitle

%%%%%%%%%%%%%%%%%%%%%%%%%%%%%%%%%%%%%%%%%%%%%%%%%%%%%%%%%%%%%%%%%%%%%%%%%%%%%
%Main text

The emergence of physical properties that are absent in the building blocks of interacting quantum systems is a central theme of condensed matter physics.
Fractionalization, a phenomenon whereby the collective excitations of a system have quantum numbers that cannot be obtained from its elementary constituents, provides powerful evidence of the \textit{more is different} paradigm~\cite{Anderson1972}.

One of the simplest models that features fractionalization is the Heisenberg Hamiltonian for spin-1 chains with nearest-neighbor antiferromagnetic interactions $\hat{\mathbf{S}}_i \cdot \hat{\mathbf{S}}_{i+1}$ (where $\hat{\mathbf{S}}_i$ denotes the vector of spin-1 operators at site $i$).
In contrast to the spin-1/2 counterparts, Haldane predicted~\cite{Haldane1983,Haldane1983a} that one-dimensional (1D) spin-1 Heisenberg antiferromagnets with \textit{periodic boundary conditions} (PBCs) should have a gapped excitation spectrum, along with a singlet ground state (GS) for which the spin-spin correlation function 
%$\langle \hat{\mathbf{S}}_i \cdot \hat{\mathbf{S}}_{i+d} \rangle$ 
decays exponentially.
%with $d$.

The seminal work of Haldane was later backed up by the analytical valence bond solid solution of the Affleck--Kennedy--Lieb--Tasaki (AKLT) model~\cite{Affleck1987}, where Haldane's conjecture was 
%rigorously 
verified for the 1D antiferromagnetic spin-1 Heisenberg Hamiltonian with additional nearest-neighbor biquadratic terms $\beta ( \hat{\mathbf{S}}_i \cdot \hat{\mathbf{S}}_{i+1} )^2$, taking $\beta=1/3$.
Remarkably, the AKLT solution for \textit{open boundary conditions} (OBCs) revealed a fourfold degenerate GS due to the emergence of two fractional spin-1/2 degrees of freedom at the chain edges.

Further numerical work~\cite{Kennedy1990,White1993a} established that the generalization of the AKLT Hamiltonian for arbitrary $\beta$, hereafter referred to as the bilinear-biquadratic (BLBQ) model, features both gapped excitations and fractional spin-1/2 edge states, with a fourfold degeneracy in the thermodynamic limit, in a range of $\beta$ that includes $0 \leq \beta \leq 1/3$.
We refer to this as the Haldane phase.

The connection between the nature of the boundary conditions and the GS degeneracy is a hallmark of topological order.
In the case of spin-1 Haldane chains, it has been shown that a hidden form of topological antiferromagnetic order, akin to that of the fractional quantum Hall effect~\cite{Girvin1989} and intimately related to a symmetry breaking~\cite{Kennedy1992}, can be detected through nonlocal string order parameters~\cite{Nijs1989,Pollmann2012a}.
It is now understood that the Haldane phase is a symmetry-protected topological phase~\cite{Gu2009,Pollmann2010}.

In this Letter, we present a Hubbard model that captures the Haldane physics of spin-1 BLBQ chains.
We consider a Hubbard Hamiltonian at half filling, defined in a 1D lattice with four sites per unit cell, and show, by numerical means, that its low-energy properties are well described by a BLBQ model in the Haldane phase, thereby connecting two celebrated milestones~\cite{Affleck1989,Tasaki1998,Arovas2021} of the last half century.
Furthermore, we find that our proposed Hamiltonian interpolates between a spin-1 Haldane chain for $U>0$ (where $U$ denotes the on-site Hubbard repulsion) and a 1D topological insulator in the noninteracting limit $U=0$.

Early work showed that the Haldane phase can be obtained in spin-1/2 Heisenberg alternating chains~\cite{Hida1992} and ladders~\cite{Barnes1993,White1996} with all couplings antiferromagnetic.
%, as opposed to more natural implementations that feature ferromagnetic interactions~\cite{Watanabe1994}.
Given that, in the limit of large $U$, the Hubbard model at half filling is known to map into a spin-1/2 antiferromagnetic Heisenberg Hamiltonian~\cite{Anderson1959}, this opens a way to realize the Haldane physics with Fermi-Hubbard systems~\cite{Sompet2021}.
It must also be noted that, by turning off the interactions of a variety of fermionic models in the Haldane phase, both trivial~\cite{Anfuso2007,Moudgalya2015} and topological~\cite{Verresen2017,Lisandrini2017,Zirnbauer2021} 1D band insulators have been adiabatically retrieved.
%, a fact that is indicative of the complex effect of interactions in the topological classification of 1D systems~\cite{Turner2011}.
In contrast to the aforementioned literature, here we put forward a Hubbard Hamiltonian that exhibits fractional spin-1/2 edge states supported in four-site fermionic structures that are robust $s=1$ spins, effectively generated for any $U>0$ without the need for additional terms.

\begin{figure}
 \includegraphics[width=\columnwidth]{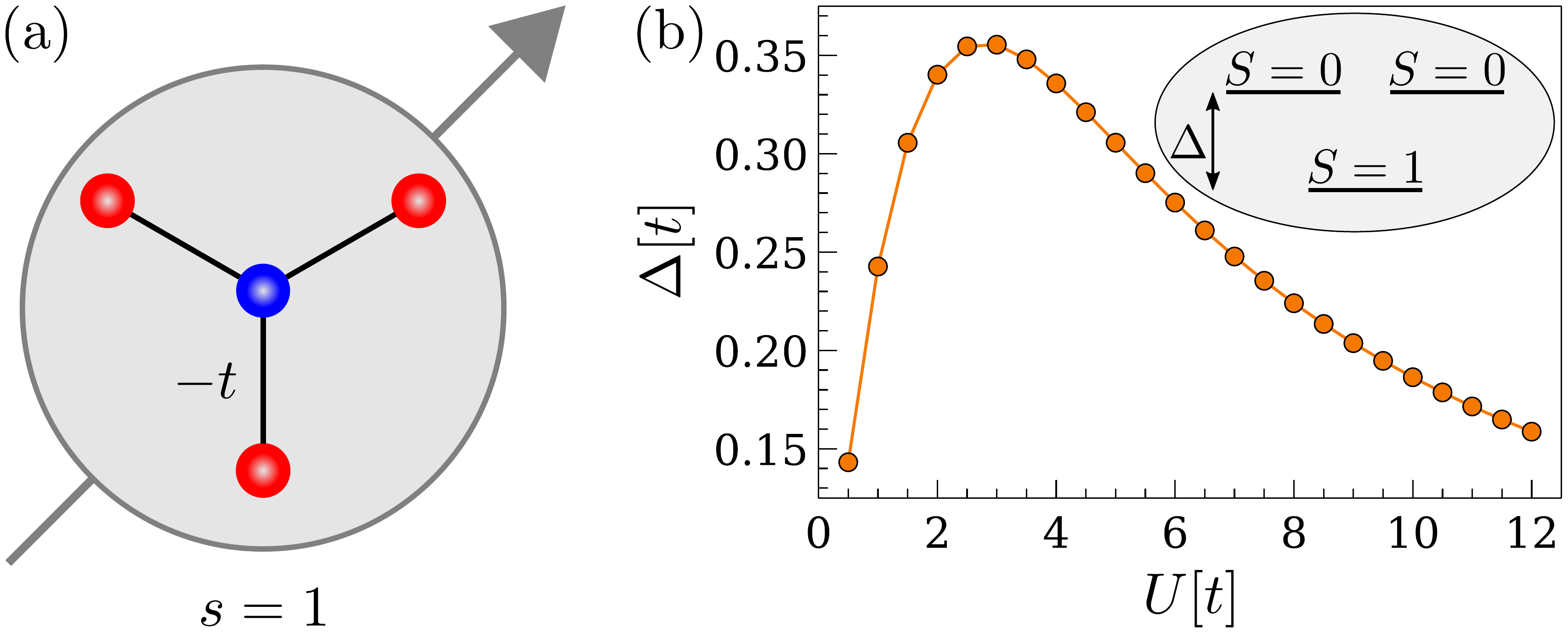}
 \caption{
 Four-site cluster as a robust spin-1 building block.
 (a) Representation of a four-site cluster
 %, with intracluster hopping $t$, 
 as an $s=1$ spin.
 Colors of small circles denote the two sublattices.
 (b) Energy splitting between the 
 %total spin 
 $S=1$ GS and the first excited state, as a function of 
 %the on-site Hubbard repulsion, 
 $U$, obtained with the Hubbard model for the four-site cluster at half filling.
 }
 \label{fig:4-site-spin1}
\end{figure}

The first key step of our strategy to realize the Haldane physics with a Hubbard model is to generate a robust $s=1$ spin.
Inspired by previous work~\cite{Ortiz2019,Mishra2021}, we propose the \textit{four-site cluster}---a fermionic structure with trigonal planar geometry in which the three outer sites are linked to the central one via an intracluster hopping $t>0$ (Fig.~\ref{fig:4-site-spin1}a).
The four-site cluster is bipartite and has a sublattice imbalance of $|N_A - N_B| = 2$ (where $N_{A/B}$ denotes the number of sites of the $A/B$ sublattice) which implies, by virtue of Lieb's theorem~\cite{Lieb1989}, that the corresponding Hubbard model at half filling must yield a GS with total spin $S=1$.
Moreover, exact diagonalization of the model yields a splitting $\Delta$ between the $S=1$ GS and a doublet of $S=0$ first excited states (Fig.~\ref{fig:4-site-spin1}b) that is at least one order of magnitude larger than the effective exchange coupling $J$ computed below.
Previous work has shown that this splitting is maximal, compared to other $|N_A-N_B|=2$ structures, due to its $\mathcal{C}_3$ symmetry~\cite{Ortiz2019}.
Therefore, we establish the four-site cluster as our robust spin-1 building block, mapping four fermionic sites into one $s=1$ spin.

\begin{figure}
 \includegraphics[width=\columnwidth]{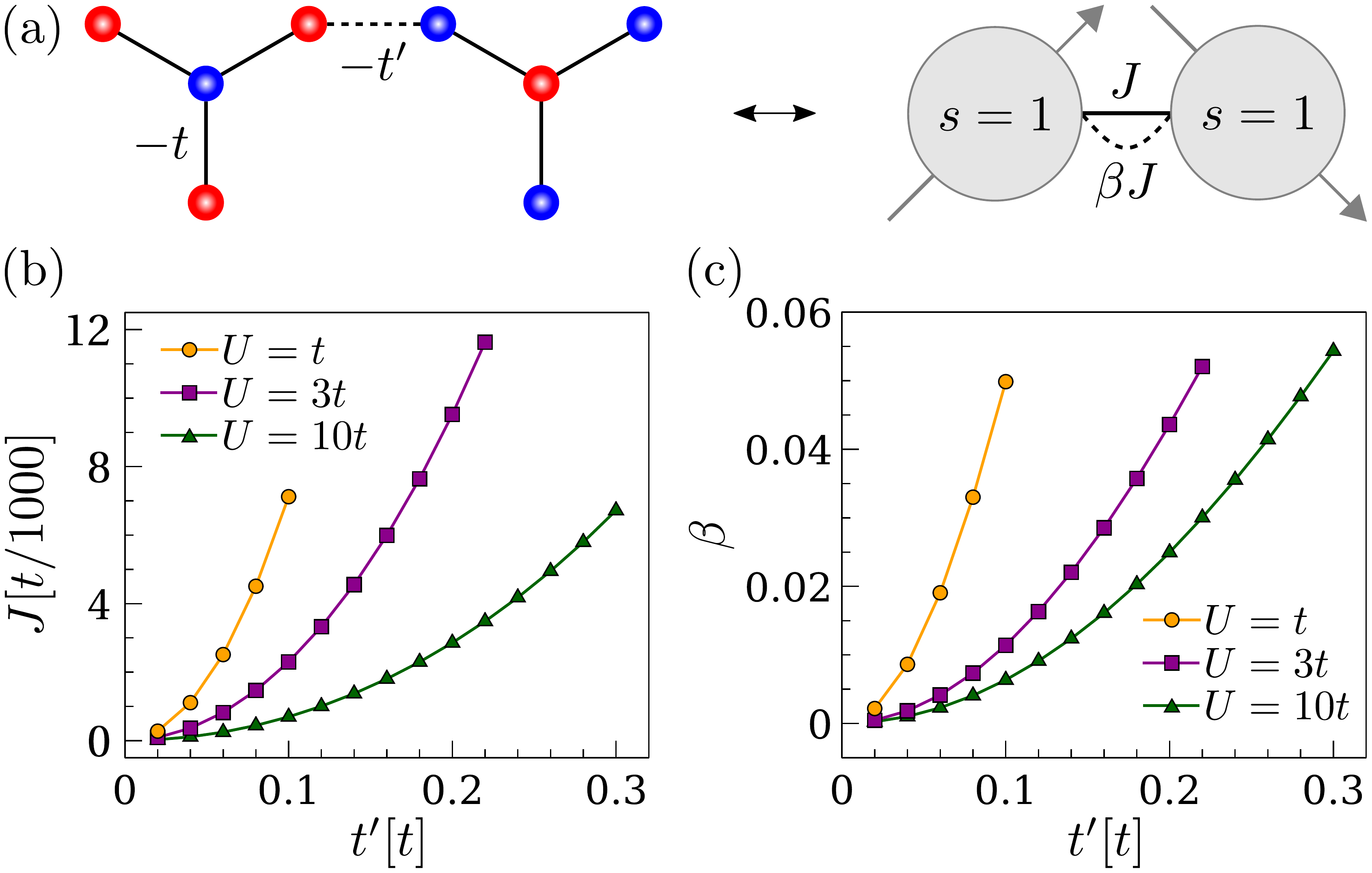}
 \caption{
 Effective spin picture of two coupled four-site clusters.
 (a) Depiction of a four-site dimer
 %, with intercluster hopping $t'$, 
 as an antiferromagnetic spin-1 dimer with bilinear ($J$) and biquadratic ($\beta J$) exchange couplings.
 (b,c) BLBQ model parameters $J$ (b) and $\beta$ (c), as a function of the Hubbard model parameters, obtained by matching the energy levels of the spin-1 BLBQ dimer to the low-energy spectrum of the four-site Hubbard dimer at half filling.
 }
 \label{fig:dimer-BLBQ-mapping}
\end{figure}

The next step toward the realization of the Haldane phase is to create a 1D lattice of antiferromagnetically coupled $s=1$ four-site clusters.
We first consider the case of $N=2$ chains (where $N$ denotes either the number of four-site clusters or the number of $s=1$ spins), which we refer to as \textit{dimers}.
To achieve an effective antiferromagnetic coupling in the four-site dimer, we introduce an intercluster hopping $t'>0$ between the closest sites of 
%each of 
the two four-site clusters (Fig.~\ref{fig:dimer-BLBQ-mapping}a).
We find that, for small enough $t'$---which preserves the picture of robust spin-1 building blocks (see the Supplemental Material for technical details on how the upper bounds of $t'$ are set)---, the four-site Hubbard dimer at half filling yields a low-energy spectrum that can be matched, spin degeneracies included, to that of the BLBQ model with an effective antiferromagnetic exchange coupling $J>0$,
\begin{equation}
\hat{\mathcal{H}}_{\text{BLBQ}} = J \sum_{i=1}^{N-1} \left[ \hat{\bm{S}}_i \cdot \hat{\bm{S}}_{i+1} + \beta \left( \hat{\bm{S}}_i \cdot \hat{\bm{S}}_{i+1} \right)^2 \right], 
\end{equation}
for $N=2$, thereby establishing a mapping between four-site Hubbard and spin-1 BLBQ dimers (Fig.~\ref{fig:dimer-BLBQ-mapping}a).
It must be noted that a small but nonzero biquadratic exchange coupling---the simplest term that can be added to $N=2$ Heisenberg chains, compatible with all the symmetries of the system---is necessary to obtain an exact match.

Figure~\ref{fig:dimer-BLBQ-mapping}b,c shows the BLBQ model parameters $J$ (Fig.~\ref{fig:dimer-BLBQ-mapping}b) and $\beta$ (Fig.~\ref{fig:dimer-BLBQ-mapping}c), 
%as a function of the Hubbard model parameters, 
obtained by fitting the energy levels of the spin-1 BLBQ dimer, solved analytically, to the low-energy spectrum of the four-site Hubbard dimer at half filling, solved by exact diagonalization.
We highlight three main features.
First, we note that $J$ is at least one order of magnitude smaller than $\Delta$, as previously anticipated.
This ascertains that the four-site clusters, albeit coupling antiferromagnetically, have a large 
%effective 
intracluster ferromagnetic exchange that preserves their spin-1 character~\footnote{A similar picture has been reported in triangulene dimers~\cite{Mishra2020}, where each triangulene plays the role of the effective $s=1$ spin.}.
Second, we verify that $J$ follows the superexchange scaling~\cite{Anderson1959}  $J = \gamma \frac{t'^2}{U}$, with $\gamma \simeq 0.716$ obtained numerically.
Third, we obtain values of $\beta$ between 0 and 0.06, in which case the BLBQ model is known to remain in the Haldane phase~\cite{Kennedy1990,White1993a}.

\begin{figure*}
 \includegraphics[width=2\columnwidth]{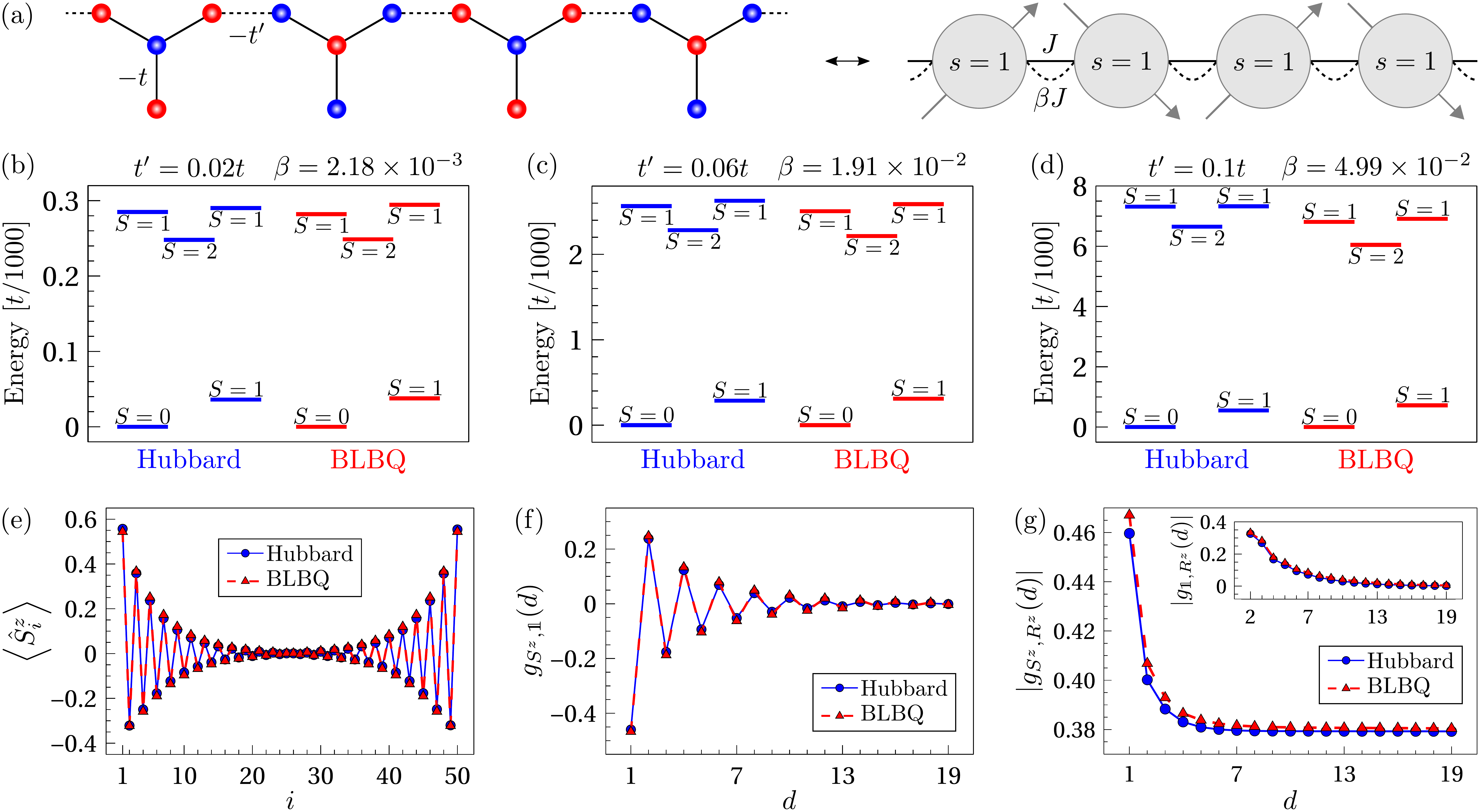}
 \caption{
 Agreement of the effective spin-1 model for 1D four-site Hubbard lattices and observation of Haldane phase properties.
 (a) Sketch of the general mapping between four-site 1D lattices and spin-1 BLBQ chains.
 (b--d) Comparison between the lowest-energy levels of $N=10$ four-site Hubbard and spin-1 BLBQ 
 %open-ended
 chains.
 Hubbard model results were obtained at half filling, for $U=t$, with $t' = 0.02 t$ (b), $t' = 0.06 t$ (c), and $t' = 0.1 t$ (d).
 BLBQ model parameters $J=2.76 \times 10^{-4} t$, $\beta=2.18 \times 10^{-3}$ (b), $J=2.51 \times 10^{-3} t$, $\beta=1.91 \times 10^{-2}$ (c), and $J=7.12 \times 10^{-3} t$, $\beta=4.99 \times 10^{-2}$ (d) were fixed by matching the low-energy spectra of the dimers.
 Energy levels were shifted as to set the GS energies to zero.
 (e--g) Average magnetization (e), spin-spin correlation function (f), and string order parameters (g), obtained for the lowest-energy state with $|S,S_z\rangle = |1,+1\rangle$ of $N=50$ four-site Hubbard and spin-1 BLBQ 
 %open-ended
 chains, using the same model parameters as in (c).
 }
 \label{fig:Hubb-Haldane-agreement}
\end{figure*}

To provide further evidence that our proposed 1D four-site Hubbard lattice at half filling captures the Haldane physics of spin-1 BLBQ chains, we now extend the comparison between both models for $N>2$ chains (Fig.~\ref{fig:Hubb-Haldane-agreement}a), keeping the same $J$ and $\beta$ as obtained for the dimer.
In what follows, we shall always consider OBCs.
We use the QuSpin package~\cite{Weinberg2017,Weinberg2019} to perform exact diagonalization on both $N=3$ four-site Hubbard lattices and spin-1 BLBQ chains up to $N=19$ (the limit within our computational resources), and the ITensor library~\cite{ITensor} for density matrix renormalization group~\cite{White1992,White1993} calculations on $N \geq 20$ spin-1 BLBQ chains and $N \geq 4$ four-site Hubbard lattices.

Figure~\ref{fig:Hubb-Haldane-agreement}b-d shows the five lowest-energy levels obtained with both Hubbard and BLBQ models for $N=10$ chains, using different sets of model parameters.
An overall good agreement is apparent, with differences verified to be smaller than $J/10$ for all the energy levels.
We also observe that, whereas the pattern of spin degeneracies is the same no matter the model parameters taken, the agreement between the energies is worse when $\beta$ is larger.
This suggests that higher-order corrections, such as bicubic exchange couplings, should be included if a better quantitative description is sought.
Finally, we observe a singlet-triplet splitting between the GS and the first excited state, verified to decay exponentially with $N$ (see the Supplemental Material), which leads to a fourfold degeneracy for sufficiently long chains.

Now that the low-energy spectra of both Hubbard and BLBQ models were shown to hold a good agreement, we move on to the comparison of other wave function properties.
For that matter, we define the correlator
\begin{equation}
 g_{\mathcal{O},\mathcal{U}} (d) = \left\langle \hat{\mathcal{O}}_p \left( \prod_{q=p+1}^{p+d-1} \hat{\mathcal{U}}_q \right) \hat{\mathcal{O}}_{p+d} \right\rangle,
\end{equation}
where $\hat{\mathcal{O}}_p$ and $\hat{\mathcal{U}}_p$ are arbitrary local operators that act on site $p$, assumed to be sufficiently away from the edges, so that $g_{\mathcal{O},\mathcal{U}} (d)$ is independent of $p$.
%and $\langle ... \rangle$ denotes the expectation value within a given state.
Notably, the spin-string correlator $|g_{S^z,R^z} (d)|$, where $\hat{R}^z_p = \mathrm{exp} ( \mathrm{i} \pi \hat{S}^z_p )$, is known to be an order parameter 
%of the Haldane phase 
that, when evaluated for a state in the fourfold GS manifold of spin-1 Haldane chains, converges to a finite value~\cite{Nijs1989} ($|g_{S^z,R^z} (\infty)| \simeq 0.37$ in the case of the Heisenberg model~\cite{White1993a}).
On the other hand, the pure-string correlator $|g_{\mathds{1},R^z} (d)|$ has been established as an order parameter 
%of the Haldane phase 
that, for the same state, exhibits the opposite behavior, vanishing in the $d \gg 1$ limit~\cite{Pollmann2012a}.
For comparison purposes, given that each four-site cluster is mapped into an $s=1$ spin, in the case of the Hubbard model we take the local spin operators as the sum of fermionic spin operators over the corresponding four-site clusters, i.e.,
\begin{equation}
 \hat{\bm{S}}_i = \frac{1}{2} \sum_{j\in \mathrm{cluster}_i}
 \begin{pmatrix}
 \hat{c}^\dagger_{j,\uparrow} & \hat{c}^\dagger_{j,\downarrow} \\
 \end{pmatrix}
 \cdot \bm{\tau} \cdot
 \begin{pmatrix}
 \hat{c}_{j,\uparrow} \\
 \hat{c}_{j,\downarrow} \\
 \end{pmatrix},
\end{equation}
where $\bm{\tau}$ is the vector of Pauli matrices and $c^\dagger_{j,\sigma}$ ($\hat{c}_{j,\sigma}$) is the creation (annihilation) operator for an electron in site $j$ with spin $\sigma=\uparrow,\downarrow$.

\begin{figure*}
 \includegraphics[width=2\columnwidth]{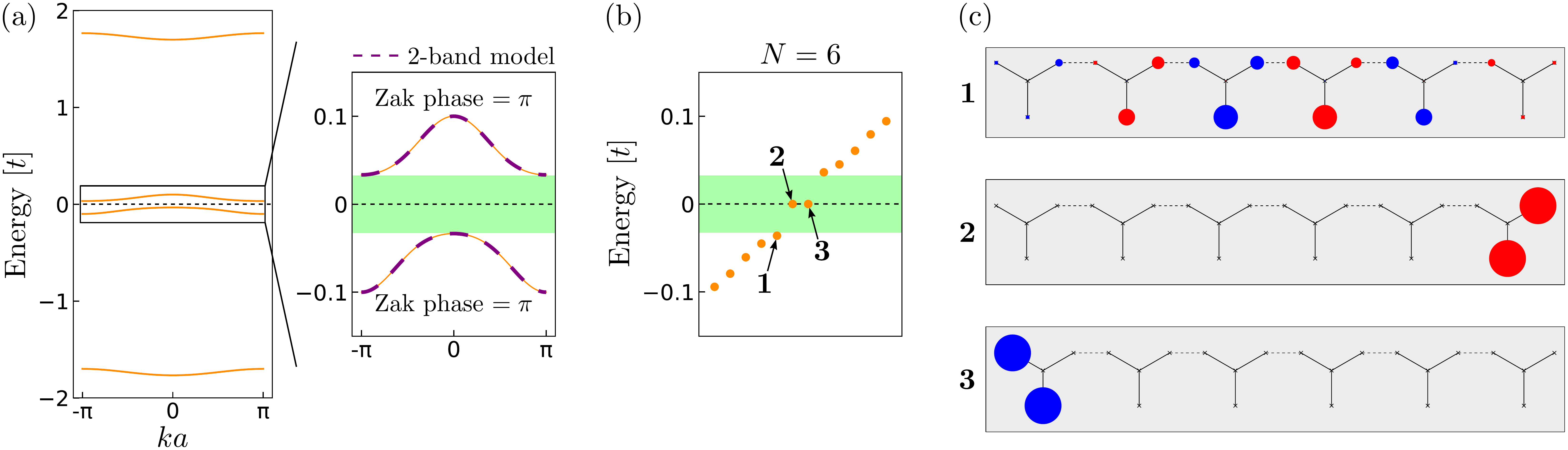}
 \caption{
 Topology of the four-site 1D lattice in the noninteracting limit.
 (a) Energy bands of the four-site 1D crystal, obtained 
 %with the tight-binding model 
 for $t' = 0.1 t$.
 The horizontal dashed line marks the Fermi level at half filling.
 The low-energy bands feature a gap (green region), along with a topological Zak phase of $\pi$, and are well described by an effective two-band model derived for $t' \ll t$.
 (b,c) Low-energy levels (b) and atomic wave function distribution of selected eigenstates (c), obtained 
 %with the tight-binding model 
 for an $N=6$ four-site chain with $t'= 0.1t$.
 The colors and size of the circles in (c) represent the two sublattices and the wave function overlap at each site, respectively.
 Two in-gap edge states are observed (\textbf{2} and \textbf{3}).
 All other eigenstates have bulk character (\textbf{1} is shown as example).
 }
 \label{fig:topology-single-particle}
\end{figure*}

In Fig.~\ref{fig:Hubb-Haldane-agreement}e--g, we present expectation values and correlators computed for the lowest-energy state with $|S,S_z\rangle = |1,+1\rangle$ (where $S_z$ denotes the total spin projection) of $N=50$ chains, using both Hubbard and BLBQ models.
Figure~\ref{fig:Hubb-Haldane-agreement}e shows the average magnetization, which features the well-known signature~\cite{White1993} of spin-1/2 edge fractionalization.
In Fig.~\ref{fig:Hubb-Haldane-agreement}f, we plot the spin-spin correlation function $g_{S^z,\mathds{1}} (d) = \langle \hat{S}^z_p \hat{S}^z_{p+d} \rangle$, which exhibits antiferromagnetic correlations that decay exponentially in the bulk, as expected in the Haldane phase due to the absence of long-range order~\cite{Haldane1983}.
Figure~\ref{fig:Hubb-Haldane-agreement}g provides the ultimate evidence that both models capture the Haldane physics, as the string order parameters $|g_{S^z,R^z} (d)|$ and $|g_{\mathds{1},R^z} (d)|$ are shown to follow the characteristic behaviors~\cite{Nijs1989,Pollmann2012a}.
Notably, both models yield a striking agreement in all these calculations, thus supporting our claim of its equivalence.

For completeness, we have verified that the Hubbard-BLBQ mapping holds not only for $U=t$ (Fig.~\ref{fig:Hubb-Haldane-agreement}b--g), but for all $U>0$.
In the Supplemental Material, we show calculations for $U=0.1t$, relevant to connect with the noninteracting limit, and for $U=10t$, in which case the Hubbard model effectively maps into a spin-1/2 
%Heisenberg 
Hamiltonian~\cite{Anderson1959}.
It must be noted that the magnitude of $U$ determines the spin-1 robustness of the four-site clusters (quantified by $\Delta$), with the main consequence being that a maximal $\Delta$ allows for a maximal $J$, a scenario that occurs for $U \simeq 3t$ (Fig.~\ref{fig:4-site-spin1}b)~\footnote{Note that even though $J$ scales as $J \propto t'^2/U$, a smaller $U$ does not necessarily imply a larger $J$ since the upper bounds of $t'$ for which the Hubbard-BLBQ mapping remains valid also depend on $U$.}.
Besides that, in the Supplemental Material we also show that the $\mathcal{C}_3$ symmetry of the four-site clusters is not a necessary requirement for the Hubbard-BLBQ mapping to hold.

We now consider the $U=0$ limit of the four-site 1D lattice.
As in the interacting case, we take $t' \ll t$.
Figure~\ref{fig:topology-single-particle}a shows the energy bands of the four-site 1D crystal ($N \rightarrow \infty$), which feature a gap at half filling.
We find that the low-energy bands possess a Zak phase of $\pi$ (see the Supplemental Material for the derivation of an effective two-band model from which this result can be obtained analytically), known to be a topological marker for 1D insulators with inversion symmetry~\cite{Zak1989}.
Expectedly~\cite{Delplace2011}, we also find that chains with finite length host two in-gap states localized at the edges (Fig.~\ref{fig:topology-single-particle}b,c).
Interestingly, these in-gap states have rigorously zero energy regardless of $N$ (Fig.~\ref{fig:topology-single-particle}b)~\footnote{Note that for even $N$, $N_A = N_B$ and Sutherland's theorem~\cite{Sutherland1986} does not warrant the presence of zero-energy states.} and are strictly localized at two edge sites (Fig.~\ref{fig:topology-single-particle}c), in contrast, for instance, to the topological edge states of the Su--Schrieffer--Heeger model~\cite{Su1979}.
The origin of such states in our model can be traced back to the presence of two sublattice-polarized~\cite{Fernandez-Rossier2007,Ortiz2019} zero-energy states in each individual four-site unit cell.
\begin{comment}
 Every four-site unit cell hosts two zero-energy states that are supported in three sites (majority sublattice).
 It is always possible to find a linear superposition so that one of them is only supported in two out of the three sites.
 In the terminal units of a chain, this gives rise to a zero-energy state that does not get hybridized by $t'$.
\end{comment}

Finally, we conjecture a topological equivalence between the $U=0$ limit of the four-site 1D lattice at half filling, shown to describe a topological insulator, and the $U>0$ regime, that realizes the Haldane symmetry-protected topological phase, according to our numerics.
Our conjecture is based on two points.
First, in both cases the GS is unique for PBCs and degenerate for OBCs.
It must be noted that, for PBCs, the unique $S=0$ GS obtained for $U>0$ is warranted by Lieb's theorem~\cite{Lieb1989} for even $N$, and numerically found via the BLBQ mapping for odd $N$, which does not contradict Lieb's theorem since PBCs in a four-site chain with odd $N$ lead to a nonbipartite lattice.
We also note that, for OBCs, there is a degeneracy enhancement at $U=0$, given that there are six degenerate configurations to accommodate two electrons in the two zero-energy edge states.
This \textit{accidental} sixfold degeneracy is lifted for any infinitesimal $U$, leaving a fourfold GS manifold. 
\begin{comment}
 1) U=0:
 1.i) PBC -> unique GS (insulator)
 1.ii) OBC -> 6fold degenerate GS for all N>1
 
 2) U>0:
 2.i) PBC -> unique S=0 GS: warranted by Lieb's theorem for even N, numerically found (via BLBQ mapping) for odd N (and not contradicting Lieb's theorem, since PBC in odd N lead to a nonbipartite lattice)
 2.ii) OBC -> 4fold degeneracy in the thermodynamic limit
\end{comment}
Second, if the Hubbard interactions do not close the gap, it is possible to adiabatically deform the $U=0$ limit into the $U>0$ regime.
We conjecture that this is the most likely scenario.

In conclusion, we have shown that a Hubbard Hamiltonian at half filling, defined in a 1D lattice that hosts one effective $s=1$ spin at each of its four-site unit cells, realizes the Haldane phase.
The proposed strategy opens a way to engineer spin-1 Haldane chains, as well as other layouts, through a variety of physical systems that are being explored for quantum simulation of the Hubbard model, such as cold atoms~\cite{Murmann2015,Mazurenko2017}, quantum dots~\cite{Hensgens2017,Mortemousque2021}, dopant arrays~\cite{Salfi2016}, hydrogenated graphene bilayer~\cite{GarciaMartinez2019}, and, in the limit of large $U$, spin-1/2 networks~\cite{Yang2017,Yang2021}.
Our findings pave the way to new experiments, going beyond early seminal works~\cite{Buyers1986,Renard1987,Hagiwara1990}, as well as the recent observations on triangulene spin chains~\cite{Mishra2021}.
Additional interest comes from the fact that both the AKLT GS and its spin-3/2 counterpart on a honeycomb lattice are resources for measurement-based quantum computing~\cite{Wei2011,Wei2012}.

%%%%%%%%%%%%%%%%%%%%%%%%%%%%%%%%%%%%%%%%%%%%%%%%%%%%%%%%%%%%%%%%%%%%%%%%%%%%%
%Acknowledgments
\begin{acknowledgments}
We thank R. Ortiz for fruitful discussions, J. L. Lado for technical assistance, and J. C. Sancho-Garc\'{i}a for access to computer facilities.
We acknowledge financial support from FCT - Funda\c{c}\~{a}o para a Ci\^{e}ncia e a Tecnologia (Grant No. SFRH/BD/138806/2018), MINECO-Spain (Grant No. PID2019-109539GB-41) and Generalitat Valenciana (Grant No. Prometeo2017/139).
\end{acknowledgments}

%%%%%%%%%%%%%%%%%%%%%%%%%%%%%%%%%%%%%%%%%%%%%%%%%%%%%%%%%%%%%%%%%%%%%%%%%%%%%
%Supplemental Material
\widetext

\pagebreak

\begin{center}
\textbf{\large Supplemental Material for ``Hubbard model for spin-1 Haldane chains''}
\end{center}

%%%%%%%%%%%%%%%%%%%%%%%%%%%%%%%%%%%%%%%%%%%%%%%%%%%%%%%%%%%%%%%%%%%%%%%%%%%%%
\section{Validity criterion for the effective spin description of coupled four-site clusters}
\label{sec:validity_criterion}

\begin{figure}[h]
 \includegraphics[width=0.75\columnwidth]{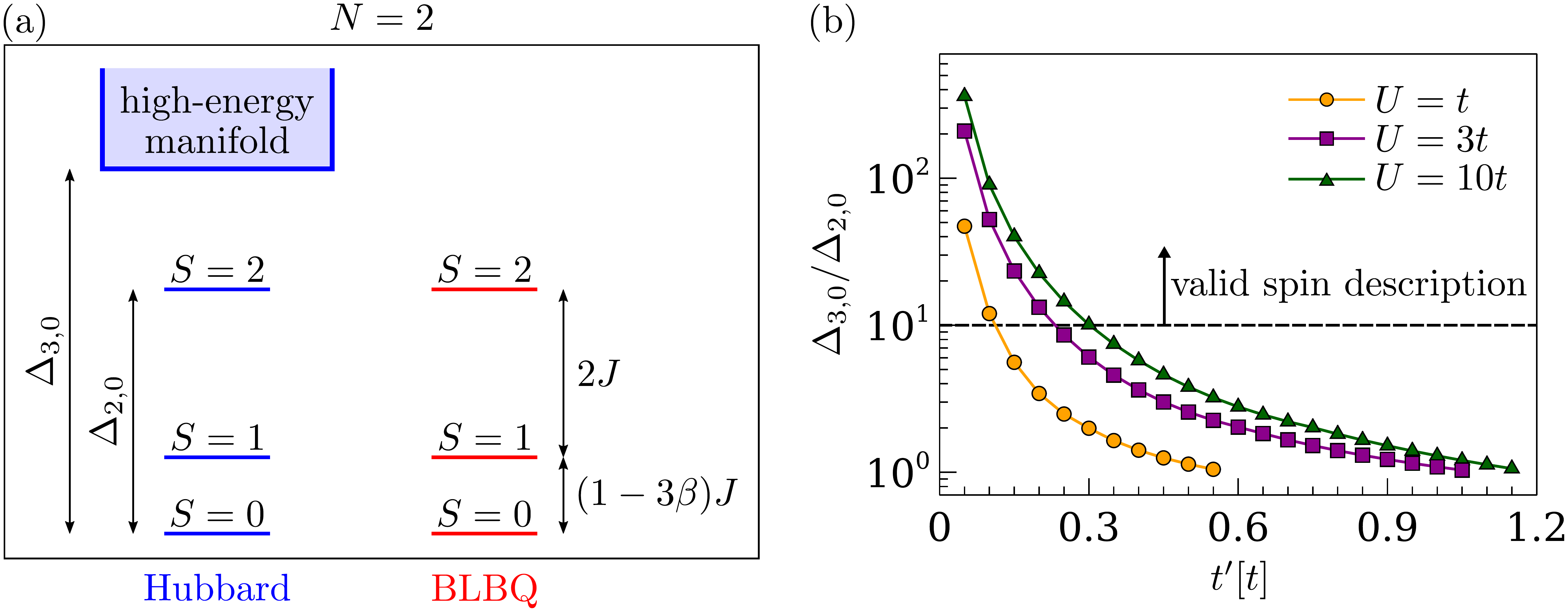}
 \caption{
 Validity criterion for the effective spin description of coupled four-site clusters.
 (a) Schematic energy level diagram of both the four-site Hubbard dimer at half filling and the spin-1 BLBQ dimer.
 (b) Separation between the high- and low-energy manifolds of the four-site Hubbard dimer at half filling, as a function of the Hubbard model parameters.
 Horizontal dashed line marks the threshold above which the effective spin description in terms of a spin-1 BLBQ Hamiltonian is deemed valid in this work.
 }
 \label{fig:SM_BLBQ-validity}
\end{figure}

In this section, we elaborate on the limitations of the effective spin description adopted in this work.
As described in the main text, we fix the parameters of the spin-1 BLBQ Hamiltonian, $J$ and $\beta$, by comparing the energy levels of the spin-1 BLBQ dimer to those of the four-site Hubbard dimer at half filling.
Given its larger Hilbert space, the four-site Hubbard dimer features more energy levels than the spin-1 BLBQ counterpart.
Therefore, to claim an accurate comparison between both spectra, we impose two conditions.
First, we require that the lowest-energy levels obtained with the Hubbard Hamiltonian match those obtained with the BLBQ model.
We have verified that this occurs always for the configuration depicted in Fig.~\ref{fig:SM_BLBQ-validity}a, which implies $J>0$ and $0 < \beta < 1/3$.
The fact that the AKLT limit~\cite{Affleck1987} $\beta=1/3$ cannot be reached through this description can be understood as a consequence of Lieb's theorem~\cite{Lieb1989}, given that it imposes a \textit{unique} $S=0$ ground state for the four-site Hubbard dimer.
Second, we require that the high-energy manifold of the four-site Hubbard dimer is separated by at least one order of magnitude with respect to the low-energy manifold in which the matching between both models occurs.
This requirement leads to restrictions on the Hubbard model parameters, as shown in Fig.~\ref{fig:SM_BLBQ-validity}b.
For instance, it is apparent that the effective spin description is always obtained for $t' \ll t$.
This result is expected since a large $t'$ would break the picture of robust $s=1$ four-site clusters.

For the sake of clarity, we point out that, in general, the Hubbard model parameters can fall into \textit{three} distinct regimes.
For simplicity, let us consider a fixed $U$ and specify the different regimes as a function of $t'$.
The first regime corresponds to values of $t'$ that are outside of the range of points plotted in Fig.~\ref{fig:SM_BLBQ-validity}b.
In such cases, $t'$ is so large that the first requirement is not achievable, i.e., the low-energy spectrum of the two model dimers cannot be matched for any $J$ and $\beta$.
In the remaining two regimes, obtained for smaller $t'$, the matching of the low-energy dimer spectra yields a BLBQ model with $J > 0$ and $0 < \beta < 1/3$, consequently in the Haldane phase~\cite{Kennedy1990,White1993a}.
However, it is observed that a larger $t'$ entails a smaller separation $\Delta_{3,0} / \Delta_{2,0}$ between the high- and low-energy manifolds (Fig.~\ref{fig:SM_BLBQ-validity}b), thus indicating a less accurate Hubbard-BLBQ mapping.
Indeed, we have verified that the increase of $\Delta_{3,0} / \Delta_{2,0}$ leads to a larger quantitative disagreement between the two models when comparing chains with $N>2$.
For that reason, we have introduced the second requirement, imposing that $\Delta_{3,0} / \Delta_{2,0} \geq 10$.
It must be noted that the threshold value imposed for this separation is arbitrary and can be modified depending on the accuracy desired.
The second (third) regime corresponds to the points that are below (above) the horizontal dashed line in Fig.~\ref{fig:SM_BLBQ-validity}b, which fail to comply (comply) with the second requirement.
In this work, we have only considered Hubbard model parameters that fall into the third regime.

\clearpage

%%%%%%%%%%%%%%%%%%%%%%%%%%%%%%%%%%%%%%%%%%%%%%%%%%%%%%%%%%%%%%%%%%%%%%%%%%%%%
\section{Singlet-triplet splitting}

A striking property commonly observed in spin-1 Haldane systems is the emergence of a low-energy singlet-triplet splitting that decays exponentially with the length in 1D chains with OBCs~\cite{Kennedy1990}.
This feature is attributed to an effective exchange coupling between the emergent fractional spin-1/2 states hosted at the chain edges.
Provided the exponential decay, a fourfold degenerate GS manifold is attained in the thermodynamic limit (or, effectively, for sufficiently long chains).

\begin{figure}[h]
 \includegraphics[width=0.45\columnwidth]{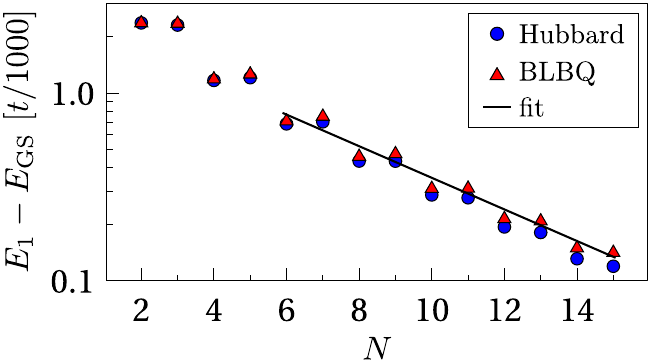}
 \caption{
 First excitation energy, as a function of the length, for 1D four-site Hubbard and spin-1 BLBQ chains.
 Hubbard model results were obtained at half filling, for $U=t$ and $t' = 0.06t$.
 BLBQ model parameters $J=2.51 \times 10^{-3} t$ and $\beta=1.91 \times 10^{-2}$ were fixed by matching the low-energy spectra of the dimers.
 Solid line indicates a fit to the function $A \mathrm{e}^{-N/\xi}$, yielding a prefactor $A \simeq 2.5 \times 10^{-3} t$ and a spin correlation length $\xi \simeq 5.2$.
 The fit was done to the BLBQ model results, considering only the chains with $N > \xi$.
 }
 \label{fig:SM_singlet-triplet-splitting}
\end{figure}

Using both the four-site Hubbard and the spin-1 BLBQ models, we have systematically found that chains with even (odd) $N$ host a singlet (triplet) GS, followed by a triplet (singlet) first excited state.
In Fig.~\ref{fig:SM_singlet-triplet-splitting}, we plot the corresponding singlet-triplet energy splitting, $E_1 - E_\mathrm{GS}$, which reveals three salient features.
First, we observe that it decays exponentially with $N$.
Second, we see a clear even-odd effect, also reported in the literature~\cite{Kennedy1990,Delgado2013,Mishra2021}.
Third, the quantitative agreement between the two models provides another strong evidence that our proposed four-site Hubbard Hamiltonian is faithfully described, at low energies, by a spin-1 BLBQ chain.

\clearpage 

%%%%%%%%%%%%%%%%%%%%%%%%%%%%%%%%%%%%%%%%%%%%%%%%%%%%%%%%%%%%%%%%%%%%%%%%%%%%%
\section{Haldane physics in four-site 1D lattices under different regimes of Hubbard interactions}

The goal of this section is to establish that the four-site 1D lattice at half filling realizes the Haldane phase for all $U>0$.
For that matter, below we show that the Hubbard-BLBQ mapping holds in two distinct regimes of Hubbard interactions---weak and strong couplings---, which complement the results presented in the main text, obtained for $U=t$.

\subsection{Weak coupling regime}
\label{subsec:weak}

To address the weak coupling regime, we set $U=0.1t$.
Applying the validity criterion detailed in Section~\ref{sec:validity_criterion}, we find that the Hubbard-BLBQ mapping is valid for $t' \lesssim 0.012 t$.
We take $t' = 0.005 t$, which leads to $J=1.69 \times 10^{-4} t$ and $\beta = 1.01 \times 10^{-2}$.

\begin{figure}[h]
 \includegraphics[width=0.75\columnwidth]{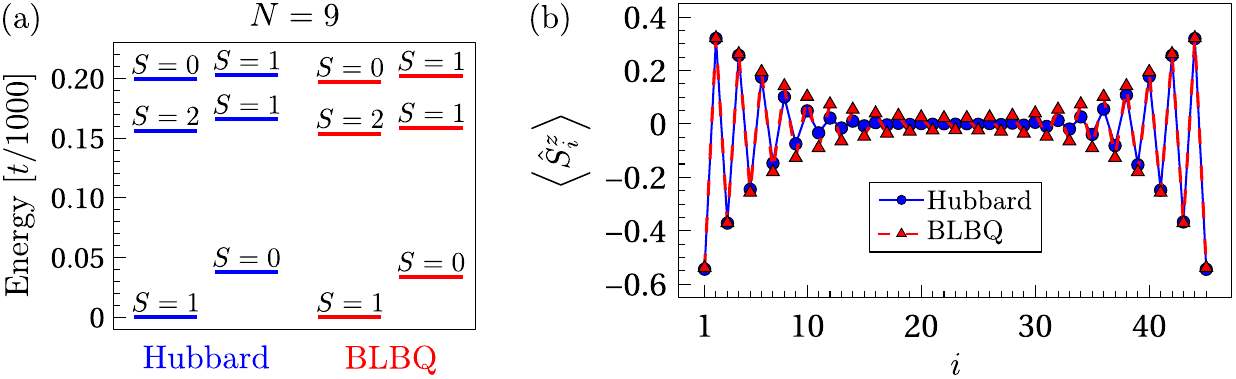}
 \caption{
 Agreement of the effective spin-1 model for four-site 1D lattices in the weak Hubbard coupling regime and observation of fractional spin-1/2 edge states.
 (a) Comparison between the lowest-energy levels of $N=9$ four-site Hubbard and spin-1 BLBQ chains.
 Hubbard model results were obtained at half filling, for $U=0.1t$ and $t' = 0.005t$.
 BLBQ model parameters $J=1.69 \times 10^{-4} t$ and $\beta = 1.01 \times 10^{-2}$ were fixed by matching the low-energy spectra of the dimers.
 Energy levels were shifted as to set the ground state energies to zero.
 (b) Average magnetization, obtained for the lowest-energy state with $|S,S_z\rangle = |1,-1\rangle$ of $N=45$ four-site Hubbard and spin-1 BLBQ chains, using the same model parameters as in (a).
 }
 \label{fig:SM_weak-coupling}
\end{figure}

In Fig.~\ref{fig:SM_weak-coupling}a, we compare the six lowest-energy levels obtained with both Hubbard and BLBQ models for $N=9$ chains.
An excellent agreement is apparent.
Figure~\ref{fig:SM_weak-coupling}b shows the average magnetization, obtained for the lowest-energy state with $|S,S_z \rangle = |1,-1 \rangle$ of $N=45$ four-site Hubbard and spin-1 BLBQ chains.
Despite some minor disagreement, most notable in the bulk sites, we observe that bold models yield the well-known signature~\cite{White1993a} of fractional spin-1/2 edge states, characteristic of the Haldane phase.

\subsection{Strong coupling limit}

We now consider the strong coupling limit, in which case the Hubbard model at half filling is known to map into an antiferromagnetic spin-1/2 Heisenberg Hamiltonian~\cite{Anderson1959}.
This means that the calculations obtained in this limit can be retrieved by considering a Heisenberg model on a four-site 1D lattice of $s=1/2$ spins, with intracluster and intercluster antiferromagnetic exchange couplings given by $J = 4t^2/U$ and $J' = 4 t'^2/U$, respectively.
To address this limit, we set $U=10t$.
The validity criterion employed in this work (see Section~\ref{sec:validity_criterion}) imposes that $t' \lesssim 0.3 t$.
We take $t' = 0.12 t$, which leads to $J=1.00 \times 10^{-3} t$ and $\beta = 9.12 \times 10^{-3}$.

\begin{figure}[h]
 \includegraphics[width=0.75\columnwidth]{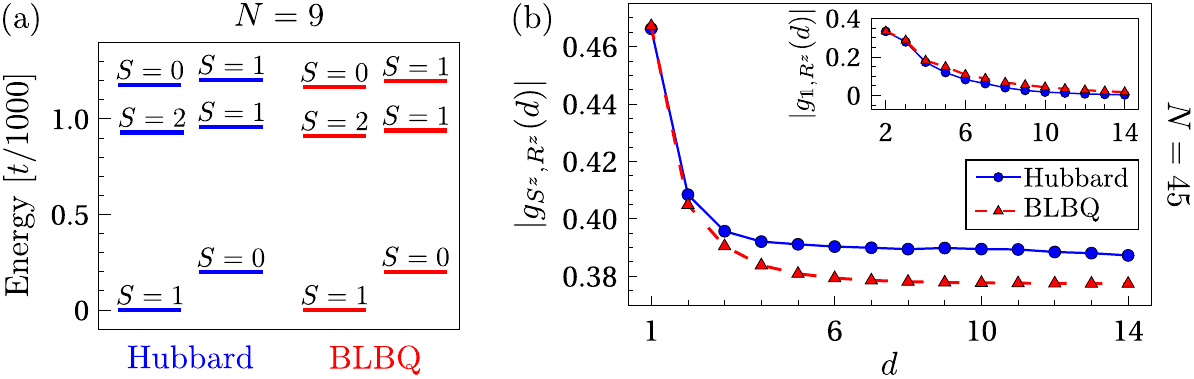}
 \caption{
 Agreement of the effective spin-1 model for four-site 1D lattices in the strong Hubbard coupling limit and observation of string order parameters in the Haldane phase.
 (a) Comparison between the lowest-energy levels of $N=9$ four-site Hubbard and spin-1 BLBQ chains.
 Hubbard model results were obtained at half filling, for $U=10t$ and $t' = 0.12t$.
 BLBQ model parameters $J=1.00 \times 10^{-3} t$ and $\beta = 9.12 \times 10^{-3}$ were fixed by matching the low-energy spectra of the dimers.
 Energy levels were shifted as to set the ground state energies to zero.
 (b) String order parameters, obtained for the lowest-energy state with $|S,S_z\rangle = |0,0\rangle$ of $N=45$ four-site Hubbard and spin-1 BLBQ chains, using the same model parameters as in (a).
 }
 \label{fig:SM_strong-coupling}
\end{figure}

Figure~\ref{fig:SM_strong-coupling}a shows the six lowest-energy levels obtained with both Hubbard and BLBQ models for $N=9$ chains, which reveal an excellent agreement.
In Fig.~\ref{fig:SM_strong-coupling}b, we plot the spin-string $|g_{S^z,R^z}|$ and the pure-string $|g_{\mathds{1},R^z}|$ correlators, calculated for the lowest-energy state with $|S,S_z\rangle = |0,0\rangle$ of $N=45$ four-site Hubbard and spin-1 BLBQ chains.
Despite some small but noticeable mismatch between both models, these string order parameters are shown to follow the characteristic behaviors of the Haldane phase~\cite{Nijs1989,Pollmann2012a}.

\clearpage

%%%%%%%%%%%%%%%%%%%%%%%%%%%%%%%%%%%%%%%%%%%%%%%%%%%%%%%%%%%%%%%%%%%%%%%%%%%%%
\section{Haldane phase for distorted four-site Hubbard chains}

In this section, we show that the Haldane physics obtained for the 1D four-site Hubbard lattice does not rely on the $\mathcal{C}_3$ symmetry of the constituent four-site clusters.
For that matter, we consider a \textit{distorted} four-site 1D lattice where this symmetry is absent and, following the same steps as in the main text, we provide numerical evidence that the corresponding Hubbard model at half filling is still well described by a spin-1 BLBQ chain in the Haldane phase.

\begin{figure}[b]
 \includegraphics[width=0.9\columnwidth]{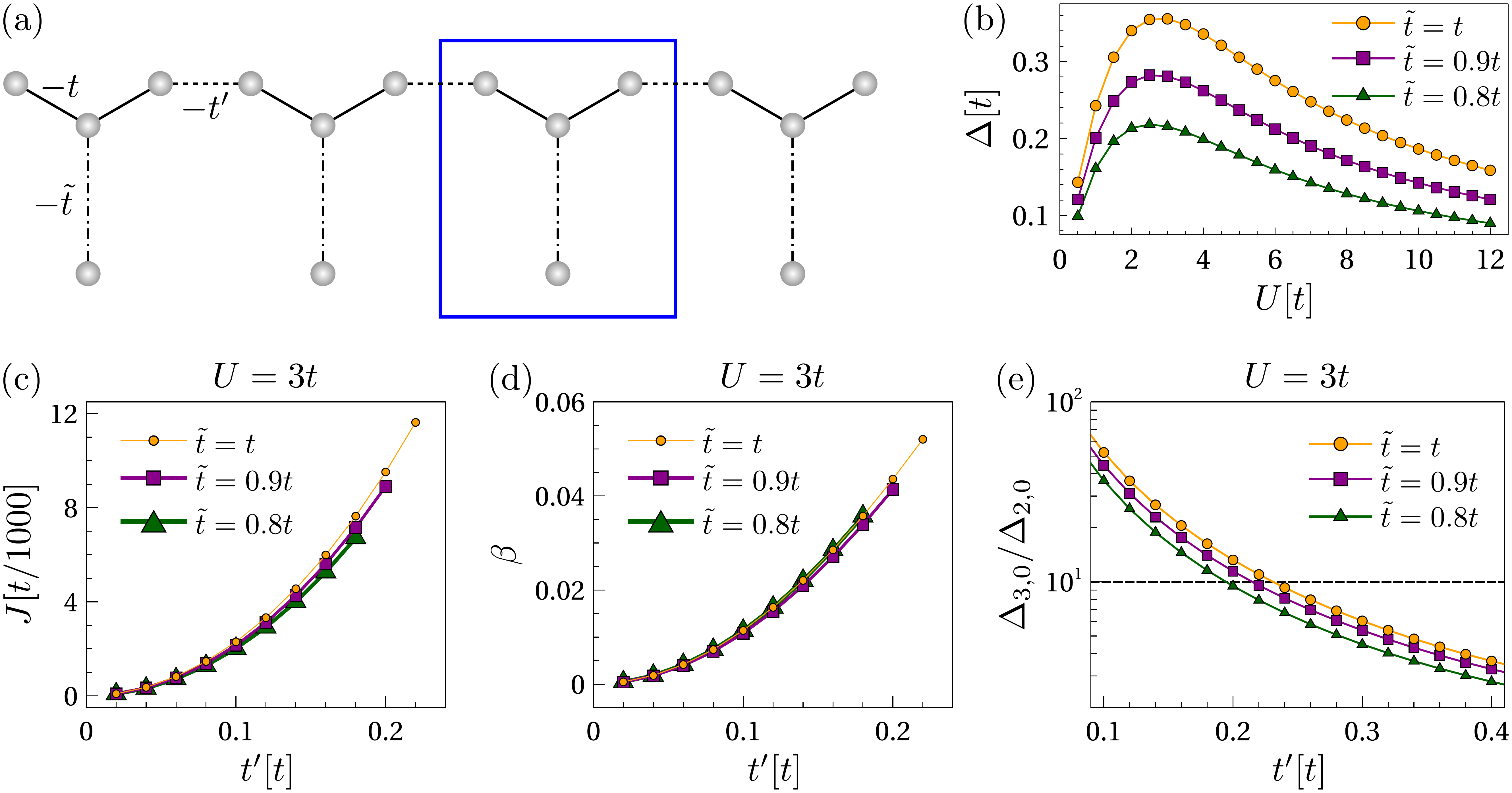}
 \caption{
 Effective spin description of the 1D distorted four-site Hubbard lattice.
 (a) Sketch of an $N=4$ distorted four-site chain.
 Blue rectangle marks a distorted four-site cluster, whose $\mathcal{C}_3$ symmetry is broken for $\tilde{t} \neq t$.
 (b) Energy splitting between the $S=1$ ground state and the $S=0$ first excited state, as a function of $U$ and $\tilde{t}$, obtained with the Hubbard model for the distorted four-site cluster at half filling.
 (c--e) BLBQ model parameters $J$ (c) and $\beta$ (d), obtained by matching the energy levels of the spin-1 BLBQ dimer to the low-energy spectrum of the distorted four-site Hubbard dimer at half filling, and separation between the low- and high-energy manifolds of the latter (e), as a function of $t'$ and $\tilde{t}$, for $U=3t$.
 Horizontal dashed line in (e) marks the threshold above which the effective spin description in terms of a spin-1 BLBQ Hamiltonian is deemed valid in this work.
 }
 \label{fig:SM_distorted-4-site}
\end{figure}

We consider a distorted four-site 1D lattice as depicted in Fig.~\ref{fig:SM_distorted-4-site}a, where the $\mathcal{C}_3$ symmetry of the four-site clusters is effectively broken by changing the hopping amplitude in the dangling bonds to $\tilde{t} \neq t$.
Whereas this symmetry breaking has no implications on the fact that the four-site Hubbard clusters have an $S=1$ ground state at half filling, as imposed by Lieb's theorem~\cite{Lieb1989}, the robustness of their spin-1 character is affected~\cite{Ortiz2019}.
Indeed, we find that, for all $U$, the splitting $\Delta$ between the $S=1$ ground state and the $S=0$ first excited state gets smaller as $\tilde{t}$ is moved away from the undistorted scenario $\tilde{t}=t$ (Fig.~\ref{fig:SM_distorted-4-site}b).
Therefore, we conclude that the absence of $\mathcal{C}_3$ symmetry is a drawback to the realization of the Haldane physics.
Nevertheless, below we shall show that it can still be attained.

In Fig.~\ref{fig:SM_distorted-4-site}c--e, we explicitly show how this $\mathcal{C}_3$ symmetry breaking affects the Hubbard-BLBQ mapping established in the case of the dimers.
Whereas moving $\tilde{t}$ away from the undistorted scenario does not lead to significant changes on $J$ (Fig.~\ref{fig:SM_distorted-4-site}c) and $\beta$ (Fig.~\ref{fig:SM_distorted-4-site}d), the fact that it implies less robust $s=1$ four-site clusters leads to stricter restrictions on the range of the Hubbard model parameters (Fig.~\ref{fig:SM_distorted-4-site}e), which in turn lead to a smaller ceiling of the BLBQ model parameters (Fig.~\ref{fig:SM_distorted-4-site}c,d).

In what follows, we consider a distorted four-site Hubbard lattice with $\tilde{t} = 0.9t$ and $U=3t$.
Using the criterion detailed in Section~\ref{sec:validity_criterion}, we find that a valid BLBQ description is obtained for $t' \leq 0.21t$.
We take $t' = 0.08t$, which leads to $J=1.37 \times 10^{-3} t$ and $\beta = 6.95 \times 10^{-3}$.

\begin{figure}
 \includegraphics[width=\columnwidth]{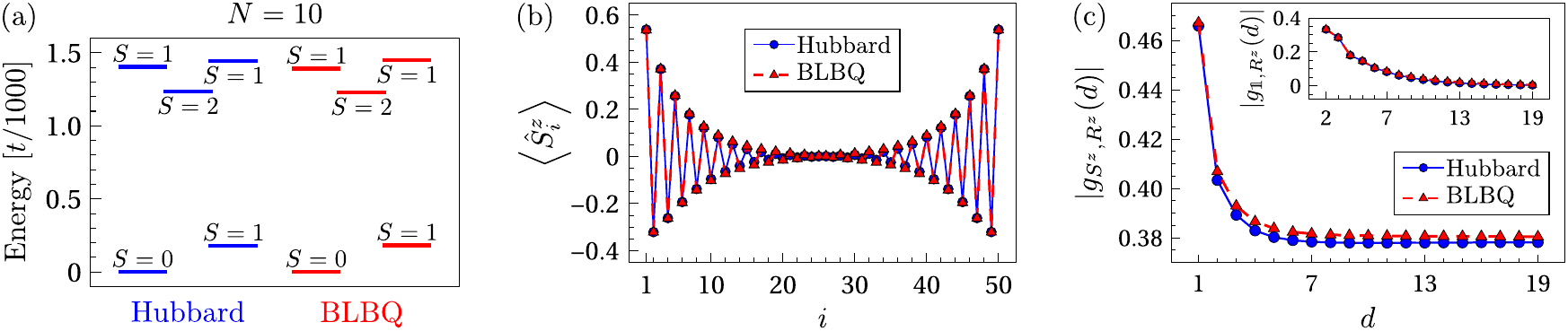}
 \caption{
 Agreement of the effective spin-1 model for distorted four-site Hubbard chains and observation of Haldane phase properties.
 (a) Comparison between the lowest-energy levels of $N=10$ distorted four-site Hubbard and spin-1 BLBQ chains.
 Hubbard model results were obtained at half filling, for $U=3t$, $\tilde{t} = 0.9t$ and $t' = 0.08t$.
 BLBQ model parameters $J=1.37 \times 10^{-3} t$ and $\beta = 6.95 \times 10^{-3}$ were fixed by matching the low-energy spectra of the dimers.
 Energy levels were shifted as to set the ground state energies to zero.
 (b,c) Average magnetization (b) and string order parameters (c), obtained for the lowest-energy state with $|S,S_z\rangle = |1,+1\rangle$ of $N=50$ distorted four-site Hubbard and spin-1 BLBQ chains, using the same model parameters as in (a).
 }
 \label{fig:SM_Haldane-distorted}
\end{figure}

Figure~\ref{fig:SM_Haldane-distorted}a shows the five lowest-energy levels obtained with both Hubbard and BLBQ models for $N=10$ chains.
An excellent agreement is apparent.
We now consider the lowest-energy state with $|S,S_z\rangle = |1,+1\rangle$ of $N=50$ chains.
In Fig.~\ref{fig:SM_Haldane-distorted}b, we plot the average magnetization, from which we observe the well-known signature~\cite{White1993a} of spin-1/2 edge fractionalization, characteristic of the Haldane phase.
Figure~\ref{fig:SM_Haldane-distorted}c features the spin-string $|g_{S^z,R^z} (d)|$ and the pure-string $|g_{\mathds{1},R^z} (d)|$ correlators, which are found to follow the characteristic behaviors of the Haldane phase~\cite{Nijs1989,Pollmann2012a}.
Notably, we observe a striking agreement between both models in these calculations.

\clearpage

%%%%%%%%%%%%%%%%%%%%%%%%%%%%%%%%%%%%%%%%%%%%%%%%%%%%%%%%%%%%%%%%%%%%%%%%%%%%%
\section{Derivation of an effective two-band model for the four-site 1D crystal}

Here we derive a two-band model that captures the low-energy electronic properties of the four-site 1D crystal at half filling.
As we shall show, this derivation is carried out for $t' \ll t$ and implies the use of a reduced Hilbert space that accounts only for the hybridization of the zero-energy states supported in each unit cell.
Below, we first derive the complete four-band model and then obtain the effective two-band Hamiltonian, whose topology is also analyzed.

\begin{figure}[b]
 \includegraphics[width=0.75\columnwidth]{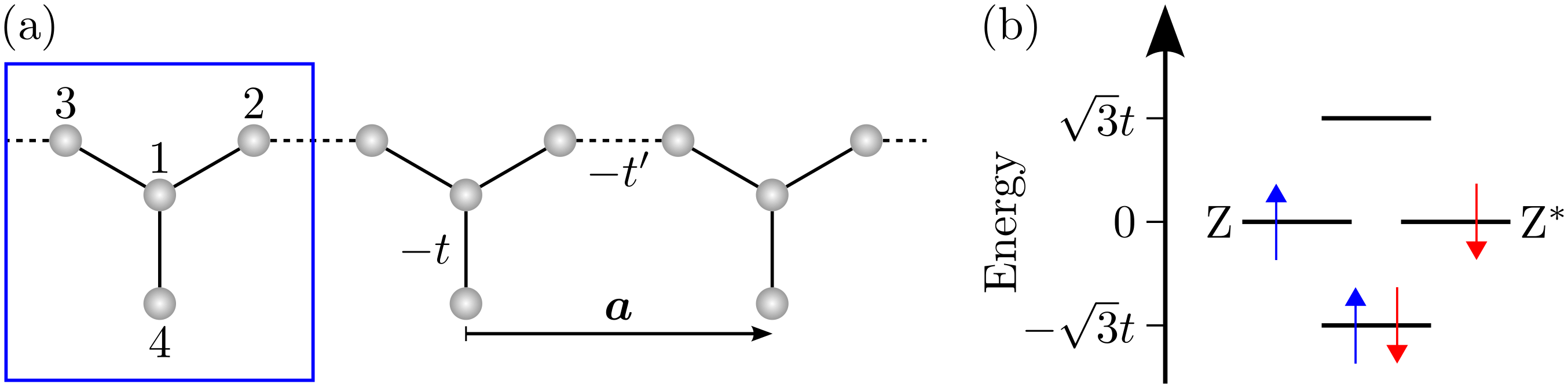}
 \caption{
 Four-site 1D crystal and diradical character of its unit cells.
 (a) Sketch of the four-site 1D crystal, where $\bm{a}$ denotes the primitive lattice vector.
 Blue rectangle marks a unit cell, whose four sites are labeled according to the numerals displayed.
 (b) Single-particle spectrum of a four-site unit cell, with the two zero-energy states Z and Z$^*$ indicated.
 Also shown is one out of the six possible electronic ground state configurations at half filling, where the blue upwards (red downwards) arrows denote electrons with spin up (down).
 }
 \label{fig:SM_4-site-1D-crystal}
\end{figure}

\subsection{Complete four-band model}

We consider the four-site 1D crystal depicted in Fig.~\ref{fig:SM_4-site-1D-crystal}a, whose unit cell sites are labeled by $j=1,2,3,4$.
The corresponding tight-binding Hamiltonian is given by
\begin{equation}
 \mathcal{H} = \sum_{\bm{R}} \left[ -t\sum_{\bar{j}=2,3,4} \left( |\bm{R},1\rangle \langle \bm{R},\bar{j}| + \mathrm{h.c.} \right) - t' \left( |\bm{R},2\rangle \langle \bm{R} + \bm{a},3| + \mathrm{h.c.} \right) \right],
 \label{eq:Hbulk}
\end{equation}
where $|\bm{R},j\rangle = |\bm{R}\rangle \otimes |j\rangle$ represents an orbital in site $j$ of a unit cell in position $\bm{R}$, $\bm{a}$ is the primitive lattice vector, and h.c. stands for hermitian conjugate.
For clarity purposes, we note that $\bm{R}$ runs over all the $N \rightarrow \infty$ unit cells in position $\bm{R} = n\bm{a}, \ n=0,1,...,N-1$.
Moreover, it must be noted that both orthonormal orbitals and PBCs are assumed.

The Hamiltonian described by Eq.~\eqref{eq:Hbulk} has translation invariance.
Therefore, Bloch's theorem ensures that it must be diagonal in reciprocal space.
To diagonalize it, we apply a Fourier transform to the spatial degree of freedom,
\begin{equation}
 |\bm{R} \rangle = \frac{1}{\sqrt{N}} \sum_{\bm{k}} \mathrm{e}^{\mathrm{i} \bm{k} \cdot \bm{R}} |\bm{k} \rangle,
\end{equation}
where, formally, the crystal momentum $\bm{k}$ runs over $\bm{k} = \frac{2\pi m}{N} \frac{\bm{a}}{a^2}, \ m=0,1,...,N-1$.
Using that
\begin{equation}
 \sum_{\bm{R}} \mathrm{e}^{\mathrm{i} \left( \bm{k} - \bm{k}' \right) \cdot \bm{R}} = N \delta_{\bm{k},\bm{k}'}, 
\end{equation}
we obtain
\begin{equation}
 \mathcal{H} = \sum_{\bm{k}} \mathcal{H}_{\bm{k}} |\bm{k} \rangle \langle \bm{k} |, 
\end{equation}
which allows to identify the Bloch Hamiltonian as
\begin{align}
 \mathcal{H}_{\bm{k}} &=  -t\sum_{\bar{j}=2,3,4} \left( |1\rangle \langle \bar{j}| + \mathrm{h.c.} \right) - t' \left( \mathrm{e}^{-\mathrm{i} ka} |2\rangle \langle 3| + \mathrm{h.c.} \right) 
 \label{eq:HBloch}
 \\ 
 &=
 \begin{pmatrix}
  |1\rangle & |2\rangle & |3\rangle & |4\rangle
 \end{pmatrix} 
 \cdot 
  \begin{pmatrix}
  0 & -t & -t & -t \\
  -t & 0 & -t' \mathrm{e}^{-\mathrm{i} ka} & 0 \\
  -t & -t' \mathrm{e}^{\mathrm{i} ka} & 0 & 0 \\
  -t & 0 & 0 & 0 
 \end{pmatrix}
 \cdot 
  \begin{pmatrix}
  \langle 1| \\
  \langle 2| \\
  \langle 3| \\
  \langle 4|
 \end{pmatrix}.
\end{align}

The above Hamiltonian has four energy bands (plotted in the main text) and contains no further assumptions besides the tight-binding approximation.

\subsection{Effective two-band Hamiltonian}

We now take the Bloch Hamiltonian obtained in Eq.~\eqref{eq:HBloch} and express it in the molecular basis.
For that matter, we use the identity
\begin{equation}
 |j \rangle = \sum_{\alpha}  \phi_{j,\alpha} |\alpha \rangle,
\end{equation}
where $\alpha$ runs over all the single-particle states $|\alpha \rangle$ of the four-site unit cell, and $\phi_{j,\alpha} = \langle \alpha |j \rangle$.
This leads to
\begin{equation}
 \mathcal{H}_{\bm{k}} = \sum_{\alpha,\alpha'} \left[ -t\sum_{\bar{j}=2,3,4} \left( \phi_{1,\alpha} \phi^*_{\bar{j},\alpha'} |\alpha\rangle \langle \alpha'| + \mathrm{h.c.} \right) - t' \left( \mathrm{e}^{-\mathrm{i} ka} \phi_{2,\alpha} \phi^*_{3,\alpha'} |\alpha\rangle \langle \alpha'| + \mathrm{h.c.} \right) \right].
 \label{eq:Hmolecular}
\end{equation}
The previous manipulation involves only a change of basis.
Therefore, Eq.~\eqref{eq:Hmolecular} is exactly equivalent to the four-band Hamiltonian.
However, by writing it in the molecular basis, it can be truncated with insight by keeping the single-particle states that are expected to capture the desired physics, in a similar spirit to the complete active space approximation~\cite{Ortiz2020}. 

At half filling, each individual four-site unit cell features two electrons occupying two zero-energy states, well separated in energy from the remaining single-particle states, which are either doubly-occupied or unoccupied (Fig.~\ref{fig:SM_4-site-1D-crystal}b).
The intercluster hopping $t'$ leads to a more convoluted picture for the four-site 1D crystal, as it promotes hybridization.
Nevertheless, for $t' \ll t$ we expect the low-energy bands at half filling to be essentially made of hybridized zero-energy states, with the remaining states acting as frozen degrees of freedom with trivial occupation.
As a consequence, the reduced Hilbert space spanned by the two zero-energy states forms a natural truncated basis for the relevant low-energy electronic properties.
In what follows, we truncate the Hamiltonian of Eq.~\eqref{eq:Hmolecular} by keeping only the zero-energy states in the sums over $\alpha,\alpha'$.

Provided the $\mathcal{C}_3$ symmetry, the wave functions of the two zero-energy states, which we label as $\alpha = \mathrm{Z},\mathrm{Z}^*$, can be written as~\cite{Ortiz2019}
\begin{equation}
 \phi_{j,\mathrm{Z}} = \frac{1}{\sqrt{3}} \left( 0, \mathrm{e}^{-\mathrm{i}\theta}, \mathrm{e}^{\mathrm{i}\theta}, 1 \right)
\end{equation}
and
\begin{equation}
 \phi_{j,\mathrm{Z}^*} = \phi^*_{j,\mathrm{Z}},
\end{equation}
with $\theta = 2\pi/3$.
After some straightforward algebra, the effective two-band Hamiltonian is obtained as
\begin{equation}
 \mathcal{H}^{\mathrm{eff}}_{\bm{k}} = -\frac{2t'}{3}
 \begin{pmatrix}
  |\mathrm{Z} \rangle & |\mathrm{Z}^*\rangle
 \end{pmatrix} 
 \cdot 
  \begin{pmatrix}
  \cos(ka - \theta) & \cos(ka) \\
  \cos(ka) & \cos(ka + \theta)
 \end{pmatrix}
 \cdot 
  \begin{pmatrix}
  \langle \mathrm{Z} | \\
  \langle \mathrm{Z}^* |
 \end{pmatrix},
\end{equation}
which from now on we write as
\begin{equation}
 \mathcal{H}^{\mathrm{eff}}_{\bm{k}} = -\frac{2t'}{3}
  \begin{pmatrix}
  \cos(ka - \theta) & \cos(ka) \\
  \cos(ka) & \cos(ka + \theta)
 \end{pmatrix},
 \label{eq:Heff}
\end{equation}
where the basis $\{ |\mathrm{Z} \rangle , |\mathrm{Z}^*\rangle \}$ is implicitly assumed.
The corresponding energy bands are given by
\begin{equation}
 E^{\mathrm{eff}}_{\bm{k},\lambda} = \frac{t'}{3} \left[ \cos(ka) + \lambda \sqrt{3+ \cos^2(ka)} \right],
\end{equation}
where $\lambda = \pm$ labels the two bands.
In the main text, we show that these bands are in excellent agreement with the low-energy bands of the complete four-band Hamiltonian.

Finally, we analyze the topology of the effective two-band model.
For that matter, we decompose the Hamiltonian of Eq.~\eqref{eq:Heff} into a linear combination of Pauli matrices $\bm{\tau}$ and the identity matrix $\mathds{1}$,
\begin{equation}
 \mathcal{H}^{\mathrm{eff}}_{\bm{k}} = \bm{d}(\bm{k}) \cdot \bm{\tau} + d_0(\bm{k}) \mathds{1},
\end{equation}
obtaining
\begin{equation}
 \bm{d}(\bm{k}) = -\frac{2t'}{3} \left(\cos(ka),0,\frac{\sqrt{3}}{2} \sin(ka)\right)
\end{equation}
and
\begin{equation}
 d_0(\bm{k}) = \frac{t'}{3} \cos(ka).
\end{equation}
We now observe that, as $\bm{k}$ runs over the first Brillouin zone, i.e. as $ka$ goes from $-\pi$ to $\pi$, the path traced out by the endpoint of $\bm{d}(\bm{k})$ is a closed ellipse on the $d_x d_z$ plane, covered counterclockwise, that encloses the origin.
Thus, we can attribute a \textit{nontrivial} winding number $\nu = 1$ to this loop, which is a signature of the Hamiltonian topology~\cite{Asboth2016}.
The winding number is intimately related to the Zak phase for 1D systems with inversion symmetry~\cite{Zak1989}, leading to a (quantized) \textit{topological} Zak phase of $\pi$ for the energy bands of $\mathcal{H}^{\mathrm{eff}}_{\bm{k}}$.

%%%%%%%%%%%%%%%%%%%%%%%%%%%%%%%%%%%%%%%%%%%%%%%%%%%%%%%%%%%%%%%%%%%%%%%%%%%%%
%Bibliography

\end{document}